\documentclass[article]{IEEEtran}
\usepackage{cite}
\usepackage{amsmath,amssymb,amsfonts}
\usepackage{algorithmic}
\usepackage{graphicx}
\usepackage{textcomp}
\usepackage{xcolor}
\usepackage{url}
\def\BibTeX{{\rm B\kern-.05em{\sc i\kern-.025em b}\kern-.08em
    T\kern-.1667em\lower.7ex\hbox{E}\kern-.125emX}}

\usepackage{subcaption}    
\usepackage[ruled,vlined]{algorithm2e}

\newtheorem{theorem}{Theorem}[section]

\def\eop{\hfill$\Box$}
\def\proof{{\em Proof: }}

\begin{document}

\title{Dither computing: a hybrid deterministic-stochastic computing framework}

\author{\IEEEauthorblockN{Chai Wah Wu}\\
\IEEEauthorblockA{IBM Research AI\\
IBM T.J. Watson Research Center\\
Yorktown Heights, NY 10598, USA \\
Email: cwwu@us.ibm.com}
}

\bibliographystyle{ieeetr}

\maketitle

\begin{abstract}
Stochastic computing has a long history as an alternative method of performing arithmetic on a computer. While it can be considered an unbiased estimator of real numbers, it has a variance and MSE on the order of $\Omega(\frac{1}{N})$. On the other hand, deterministic variants of stochastic computing remove the stochastic aspect, but cannot approximate arbitrary real numbers with arbitrary precision and are biased estimators. However, they have an asymptotically superior MSE on the order of  $O(\frac{1}{N^2})$. Recent results in deep learning with stochastic rounding suggest that the bias in the rounding can degrade performance. We proposed an alternative framework, called dither computing, that combines aspects of stochastic computing and its deterministic variants and that can perform computing with similar efficiency, is unbiased, and with a variance and MSE also on the optimal order of $\Theta(\frac{1}{N^2})$. We also show that it can be beneficial in stochastic rounding applications as well. We provide implementation details and give experimental results to comparatively show the benefits of the proposed scheme.
\end{abstract}


\section{introduction}

Stochastic computing \cite{gaines:1967,Alaghi2013,Chen2014,Duarte2015} has a long history and is an alternative framework for performing computer arithmetic using stochastic pulses. While not as efficient as binary encoding in representing numbers, it does provide a simpler implementation for arithmetic units and can tolerate errors, making it a suitable choice for computational substrates that are inherently noisy such as analog computers and quantum systems.
Stochastic computing can approximate arbitrary real numbers and perform arithmetic on them to reach the correct value in expectation, but the stochastic nature means that the result is not precise each time. Recently, Ref. \cite{Jenson2016} suggests that deterministic variants of stochastic computing can be just as efficient, and does not have the random errors introduced by the random nature of the pulses. Nevertheless in such deterministic variants the finiteness of the scheme implies that it cannot approximate general real numbers with arbitrary accuracy.
This paper proposes a framework that combines these two approaches to get the best of both worlds, and inherit some of the best properties of both schemes.
In the process, we also provide a more complete probabilistic analysis of these schemes. 
In addition to considering both the first moment of the approximation error (e.g. average error) and the variance of the representation, we also consider the set of real numbers that are represented and processed to be drawn from an independent distribution as well. This allows us to provide a more complete picture of the tradeoffs in the bias, variance of the approximation and the number of pulses along with the prior distribution of the data.

\section{Representation of real numbers via sequences}
For a real number $x$, $round(x)$ has the usual definition of $round(x) = \lfloor x+0.5\rfloor$.
We consider two independent random variables $X$, $Y$ with support in the unit interval $[0,1]$. A common assumption is that $X$ and $Y$ are uniformly distributed. The interpretation is that $X$ and $Y$ generate the real numbers that we want to perform arithmetic on. In order to represent a sample $x\in [0,1]$ from $X$ the main idea of stochastic computing (and other representations such as unary coding \cite{davis:1994}) is to use a sequence of $N$ binary pulses. In particular, $x$ is represented by a sequence of independent $N$ Bernoulli trials $X_i$. We estimate $x$ via $X_s = \frac{1}{N}\sum_{i=1}^N X_i$. Our standing assumption is that $X$, $Y$, $X_i$ and $Y_i$ are all independent. We are interested in how well $X_s$ approximates a sample $x$ in $X$. In particular, we define  $L_x =  E((X_s-x)^2|X=x)$ and are interested in the expected mean squared error (EMSE) defined as $L = E_X(L_x)$. Note that $L_x$ consists of two components, bias and variance, and the bias-variance decomposition \cite{James2013} is given by
 $L_x = Bias(X_s,x)^2 + Var(X_s)$ where $Bias(X_s,x) = E(X_s)-x$.
The following result gives a lower bound on the EMSE:

\begin{theorem}
$L \geq \frac{1}{N^2}\int_0^1 p_X(x) (Nx-\mbox{round}(Nx))^2 dx$.
\end{theorem}
\proof
Follows from the fact that $X_s$ is a rational number with denominator $N$ and thus $|X_s-x| \geq \frac{1}{N}|Nx-\mbox{round}(Nx)|$.
\eop

Given the standard assumption that $X$ is uniformly distributed, this implies that 
$L \geq 2N\int_{0}^{\frac{1}{2N}} x^2 dx = \frac{1}{12 N^2}$, i.e. the EMSE can decrease at a rate of at most $\Omega(\frac{1}{N^2})$.

In the next sections, we analyze how well $X_s$ approximates samples in $X$ asymptotically as $N\rightarrow \infty$ by analyzing the error $L$ for various variants of stochastic computing. 

\subsection{Stochastic computing} \label{sec:stochastic_computing}
A detailed survey of stochastic computing can be found in Ref. \cite{Alaghi2013}. We give here a short description of the unipolar format.  
Using the notation above, $X_i$ are chosen to be iid Bernoulli trials with
 $p(X_i = 1) = x$. 
 Then
$E(X_s) = x$ and $X_s$ is an unbiased estimator of $x$. 
Since $Bias(X_s,x) = 0$ and $Var(X_s) = \frac{1}{N}x(1-x)$, i.e. $Var(X_s) = \Omega(\frac{1}{N})$, we have $L_x = \Omega(\frac{1}{N})$ for $x\in (0,1)$.
 More specifically, if $X$ has a uniform distribution on $[0,1]$, then $L = \frac{1}{N}\int_{0}^1 x(1-x)dx = \frac{1}{6N}$.

\subsection{A deterministic variant of stochastic computing} \label{sec:deterministic_variant}
In \cite{Jenson2016} deterministic variants of stochastic computing\footnote{In the sequel for brevity we will sometimes refer to these schemes simply as ``deterministic variants''.} are proposed. Several approaches such as clock dividing, and relative prime encoding are introduced and studied. One of the benefits of a deterministic algorithm is the lack of randomness, i.e. the representation of $x$ via $X_i$ does not change and $Var(X_s) = 0$. However, the bias term $Bias(X_s,x)$ can be nonzero. Because $x$ is represented by counting the number of 1's in $X_i$, it can only represent fractions with denominator $N$. For $x = \frac{m}{2N}$ where $m$ is odd, the error is $X_s-x = \frac{1}{2N}$. This means that such values of $x$, $L_x = Bias(X_s,x)^2 + Var(X_s) = \frac{1}{4N^2} = O(\frac{1}{N^2})$. 
If $X$ is a discrete random variable with support only on the rational points $\frac{m}{N}$ for integer $0\leq m\leq N$, then $L= 0$. However, in practice, we want to the represent arbitrary real numbers in $[0,1]$. Assume that $X$ is uniformly distributed in $[0,1]$. By symmetry, we only need to analyze the $x\in [0,\frac{1}{2N}]$. 
Then $X_s-x = x$ and $L_x = x^2$. It follows that $L = 2N\int_{0}^{\frac{1}{2N}} x^2 dx = \frac{1}{12N^2} =  O(\frac{1}{N^2})$.

\subsection{Stochastic rounding} \label{sec:sround}
As the deterministic variant (Sect. \ref{sec:deterministic_variant}) has a better asymptotic EMSE than stochastic computing (Sect. \ref{sec:stochastic_computing}), one might wonder why is stochastic computing useful. It is instructive to consider a special case: 1-bit stochastic rounding \cite{Hoehfeld1992}, in which rounding a number $x\in [0,1]$ is given as a Bernoulli trial $X_1$ with $P(X_1 = 1) = x$.   
This type of rounding is equivalent to the special case $N=1$ of the stochastic computing mechanism in Sect. \ref{sec:stochastic_computing}.
In deterministic rounding, $X_1 = \mbox{round}(x)$ and the corresponding EMSE is $\tilde{L} = \frac{1}{12}$.
For stochastic rounding, $X_1$ has a Bernoulli distribution. If $P(X_1=1) = p$, then $L_x = Bias^2+Var=(p-x)^2 + p(1-p) = p(1-2x)+x^2$. Since $\frac{\partial L_x}{\partial p} = 1-2x$, it follows that for $x\in [0,\frac{1}{2}]$, $L_x$ is minimized when $p=0$ and for $x\in [\frac{1}{2},1]$, $L_x$ is minimized when $p=1$, i.e. $L_x$ is minimized when $p = \mbox{round}(x)$.
Thus $L_x \geq \tilde{L}_x$ with equality exactly when $p = \mbox{round}(x)$.
This shows that the EMSE for deterministic rounding is minimal among all stochastic rounding schemes. Thus at first glance, deterministic rounding is preferred over stochastic rounding.
While deterministic rounding has a lower EMSE than stochastic rounding, it is a biased estimator. This is problematic for application such as reduced precision deep learning where an unbiased estimator such as stochastic rounding has been shown to provide improved performance over a biased estimator such as deterministic rounding. As indicated in \cite{connolly:2020}, part of the reason is that subsequent values that are rounded deterministically are correlated and in this case the stochastic rounding prevents stagnation.

\subsection{Dither computing: A hybrid deterministic-stochastic computing framework} \label{sec:dither}
The main goal of this paper is to introduce dither computing, a hybrid deterministic-stochastic computing framework that combines the benefits of stochastic computing (Sec. \ref{sec:stochastic_computing}) and its deterministic (Sec. \ref{sec:deterministic_variant}) variants and eliminates the bias component while preserving the optimal $O(\frac{1}{N^2})$ asymptotic rate for the EMSE $L$. Furthermore, it converges to the zero bias faster than stochastic computing. The main idea here is to approximate deterministically as close as possible to the desired quantity and to stochastically approximate the remaining difference.
The encoding is constructed as follows. Let $\sigma$ be a permutation of $\{1,2,\cdots , N\}$.

For $x\in [0,\frac{1}{2}]$, let $n = \lfloor Nx \rfloor \leq \frac{N}{2}$ and $0\leq r = x - \frac{n}{N} \leq \frac{1}{N}$.
Then we pick the $N$ Bernoulli trials with $P(X_{\sigma(i)}=1) = 1$ for $1\leq i\leq n$ and $P(X_{\sigma(i)}=1) = \delta$ for $n+1\leq i \leq N$ with $\delta = \frac{Nr}{N-n}$.
Then $E(X_s) = \frac{1}{N}\left(n+\delta(N-n)\right) = x$. In addition, since $n\leq \frac{N}{2}$ and $rN\leq 1$, this implies that $\delta \leq \frac{2}{N}$. $Var(X_s) = \frac{1}{N^2}(N-n)\delta(1-\delta) \leq \frac{2}{N^2} = O(\frac{1}{N^2})$. Thus the bias is 0 and the EMSE is of the order $O(\frac{1}{N^2})$. It is clear that this remains true if $\sigma$ is either a deterministic or a random permutation as $X_s$ does not depend on $\sigma$.

For $x\in (\frac{1}{2},1]$, let $n = \lceil Nx \rceil \geq \frac{N}{2}$ and $0\leq r = \frac{n}{N}-x \leq \frac{1}{N}$.
We pick the $N$ Bernoulli trials with $P(X_{\sigma(i)}=1) = 1-\delta$ for $1\leq i\leq n$ and $P(X_{\sigma(i)}=1) = 0$ for $n+1\leq i \leq N$ with $\delta = \frac{rN}{n}$.
$E(X_s) = \frac{n(1-\delta)}{N} = x$. In addition, since $n\geq \frac{N}{2}$ and $rN\leq 1$, this implies that $\delta \leq \frac{2}{N}$. $Var(X_s) = \frac{n}{N^2}\delta(1-\delta)\leq \frac{2}{N^2}  = O(\frac{1}{N^2})$. Thus again the bias is 0 and the EMSE is of the order $O(\frac{1}{N^2})$.

The above analysis shows that dither computing offers better EMSE error than stochastic computing while preserving the zero bias property. In order for such representations to be useful in building computing machinery, we need to show that this advantage persists under arithmetic operations such as multiplication and (scaled) addition. 

\section{Multiplication of values} \label{sec:mult}
In this section, we consider whether this advantage is maintained for these schemes for multiplication of sequences via bitwise AND. The sequence corresponding to the product of $X_i$ and $Y_i$ is given by $Z_i = X_iY_i$ and the product $z=xy$ is estimated via $Z_s = \frac{1}{N}\sum_i Z_i$. 

\subsection{Stochastic computing}
In this case we want to compute the product $z = xy$.
Let $X_i$ and $Y_i$ be independent with $P(X_i=1) = x$ and $P(Y_1=1) = y$, Then for $Z_i = X_iY_i$, $Z_i$ are Bernoulli with $P(Z_i=1) = xy$ and
$E(Z_s) = xy =z$. $Var(Z_i) =z(1-z)$ and $Var(Z_s) = \frac{1}{N}z(1-z) = \Omega(\frac{1}{N})$.Thus  $Bias(Z_s,z) = 0$ and the variance and the MSE of the product maintains the suboptimal $\Omega(\frac{1}{N})$ asymptotic rate.

\subsection{Deterministic variant of stochastic computing} \label{sec:det_mult}
For numbers $x, y \in [0,1]$, we consider a unary encoding for $x$, i,e, $P(X_i = 1) = 1$ for $1\leq i\leq R$ and $P(X_i=1) = 0$ otherwise, where $R = \mbox{round}(Nx)$.
For $y$ we have $P(Y_i)=1$ if $\lfloor iy\rfloor \neq \lfloor (i+1)y \rfloor$. Let $m$ be the number of indices $i$ such that $Z_i\neq 0$, 
then $|m-yR| \leq 1$ and $Z_s = m/N$. This means that $|Z_s-z| \leq |\frac{m}{N}-\frac{yR}{N}| + |\frac{yR}{N}-xy|$. Since $|R-Nx| \leq 1$, this implies that 
$|Z_s-z|\leq \frac{2}{N}$ and
thus the bias is on the order of $O(\frac{1}{N})$ and the EMSE $L$ is on the order of $O(\frac{1}{N^2})$.

\subsection{Dither computing}
For numbers $x, y \in [0,1]$, we consider the encoding in Section  \ref{sec:dither} with the permutation $\sigma_x$ for $x$ defined as the identity and the permutation $\sigma_y$ for $y$ defined as spreading $1$ bits in a sample $(y_1,\cdots,y_N)$ of $(Y_1,\cdots Y_N)$ as much as possible.
In particular, let $y_i$ be a sample of $Y_i$ and $s_y = \sum_i y_i$.
Then $\sigma(i) = \lfloor is_y+T\rfloor \mod N$ for $i = 1,\cdots ,\lfloor\frac{N}{s_y}\rfloor$, where $T$ is a uniformly distributed random variable on [0,1] independent from $X_i$ and $Y_i$.  
We will only consider the case $x,y\in (\frac{1}{2},1]$ as the other cases are similar. Let $n_x = \lceil Nx \rceil$, $n_y = \lceil Ny \rceil$, $\delta_x = 1-\frac{Nx}{n_x}$ and $\delta_y = 1-\frac{Ny}{n_y}$.  Then $Z_i = (1-\delta_x)(1-\delta_y) =  \frac{N^2z}{n_xn_y} \neq 0$ for $\frac{n_xn_y}{N}$ of indices on average and $0$ otherwise. This implies that $E(Z_s) = z$ and the bias is $0$. Similar to the deterministic variant, it can be shown that $|Z_s-z| \leq \frac{c}{N}$ for a constant $c>0$, and thus $L$ is $O(\frac{1}{N^2})$.

\section{Scaled addition (or averaging) of values}
For $x,y\in [0,1]$, the output of the scaled addition (or averaging) operation is $u = \frac{1}{2}(x+y)\in [0,1]$. An auxiliary control sequence $W_i$ of bits is defined that is used to toggle between the two sequences by defining
$U_i$ as alternating between $X_i$ and $Y_i$: $U_i = W_iX_i + (1-W_i)Y_i$ and $u$ is estimated via  $U_s = \frac{1}{N}\sum_i U_i$.

\subsection{Stochastic computing}
The control sequence $W_i$ is defined as $N$ independent Bernoulli trials with $P(W_i = 1) = \frac{1}{2}$. It is assumed that $W_i$, $X_j$ and $Y_k$ are independent.
Then 
$E(U_s) = \frac{1}{2} E(X_s) + \frac{1}{2} E(Y_s) = \frac{1}{2}(x+y) = u$, i.e., $Bias(U_s,u) = 0$.
$Var(U_s) = \frac{1}{2N}\left(x(1-\frac{1}{2}x) +  y(1-\frac{1}{2}y)\right) = \Omega(\frac{1}{N})$.
Thus $L = \Omega(\frac{1}{N})$.

\subsection{Deterministic variant of stochastic computing}
For this case $W_i$ are deterministic and we define
$W_i = 1$ if $i$ is even and $W_i = 0$ otherwise. Let $N_e = \lfloor \frac{N}{2}\rfloor$ and $N_{o} = N-N_e$ be the number of even and odd numbers in $\{1,\cdots , N\}$ respectively.
Then $Var(E_s) = 0$ and $E(U_s) = \frac{N_e}{N} E(X_s) + \frac{N_o}{N} E(Y_s)$. If $N$ is even, $E(U_s) = \frac{1}{2}(E(X_s)+E(Y_s))$. If $N$ is odd,
$|\frac{N_e}{N}-\frac{1}{2}| - O(\frac{1}{N})$. In either case, $|E(U_s)-u| = O(\frac{1}{N})$ and $Bias(U_s,u) = O(\frac{1}{N^2})$,  $L = O(\frac{1}{N^2})$.

\subsection{Dither computing}
We set $\sigma_x$ and $\sigma_y$ both equal to the identity permutation and define sequence $\{s_i\}$ with $s_i = 1$ for $i$ odd and $0$ otherwise. With probability $\frac{1}{2}$, $W_i = s_i$ for all $i$ and $W_i = 1-s_i$ for all $i$ otherwise.
Thus the 2 sequences $\{s_i\}$ and $\{1-s_i\}$ are each chosen with probability $\frac{1}{2}$.
Note that $W_i$ and $W_j$ are correlated, $E(W_i) = \frac{1}{2}$ and $Var(W_i) = \frac{1}{4}$.
This means that $E(U_s) = u$ and the bias is $0$. The 2 sequences for $W_i$ selects 2 disjoint sets of  random variables $X_i$ and $Y_i$ the average of which has variance $O(\frac{1}{N^2})$. $U_s$ can be written as $U_s = WA_s + (1-W)B_s = B_s + W(A_s-B_s)$, where $W$ is a Bernoulli trial with $p=\frac{1}{2}$, $A_s = \frac{1}{N}\sum_{i} s_iX_i + (1-s_i)Y_i$, and
$B_s = \frac{1}{N}\sum_{i} s_iY_i + (1-s_i)X_i$. Since $A_s$ and $B_s$ takes about half of all the $X_i$ and the $Y_i$ alternating over the indices $i$, it is easy to show that $E(A_s-B_s) = O(\frac{1}{N})$. Furthermore $Var(A_s) = O(\frac{1}{N^2})$, $Var(B_s) = O(\frac{1}{N^2})$ and the formula for the variance of products of independent r.v's shows that $Var(U_s) = O(\frac{1}{N^2})$ and thus $L = O(\frac{1}{N^2})$.

\section{Numerical results}
\begin{figure}[htbp]
\centerline{\includegraphics[width=2.71in]{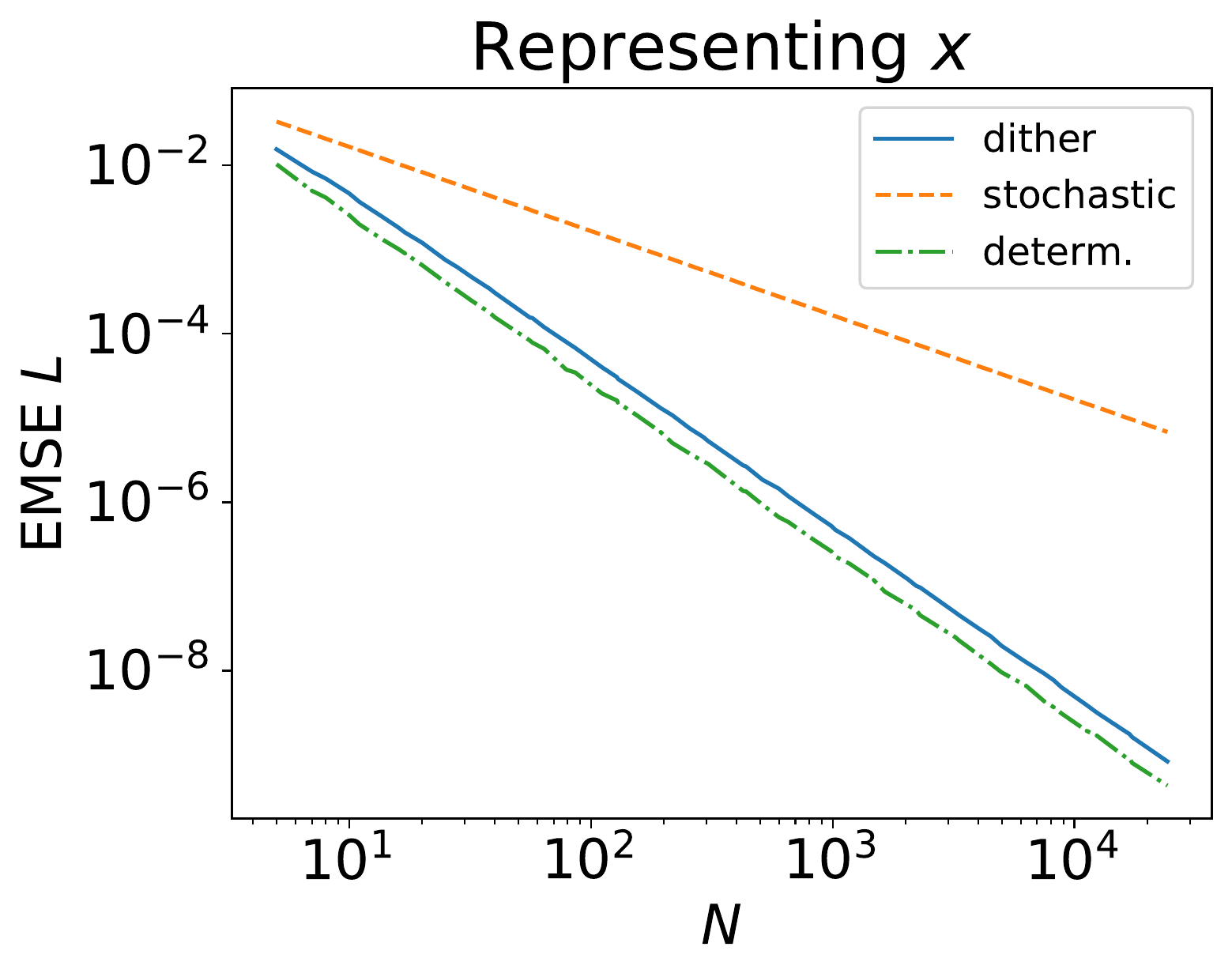}}
\caption{Sample estimate of EMSE $L$ to represent $x$ for various values of $N$.}
\label{fig:one}
\end{figure}

\begin{figure}[htbp]
\centerline{\includegraphics[width=2.71in]{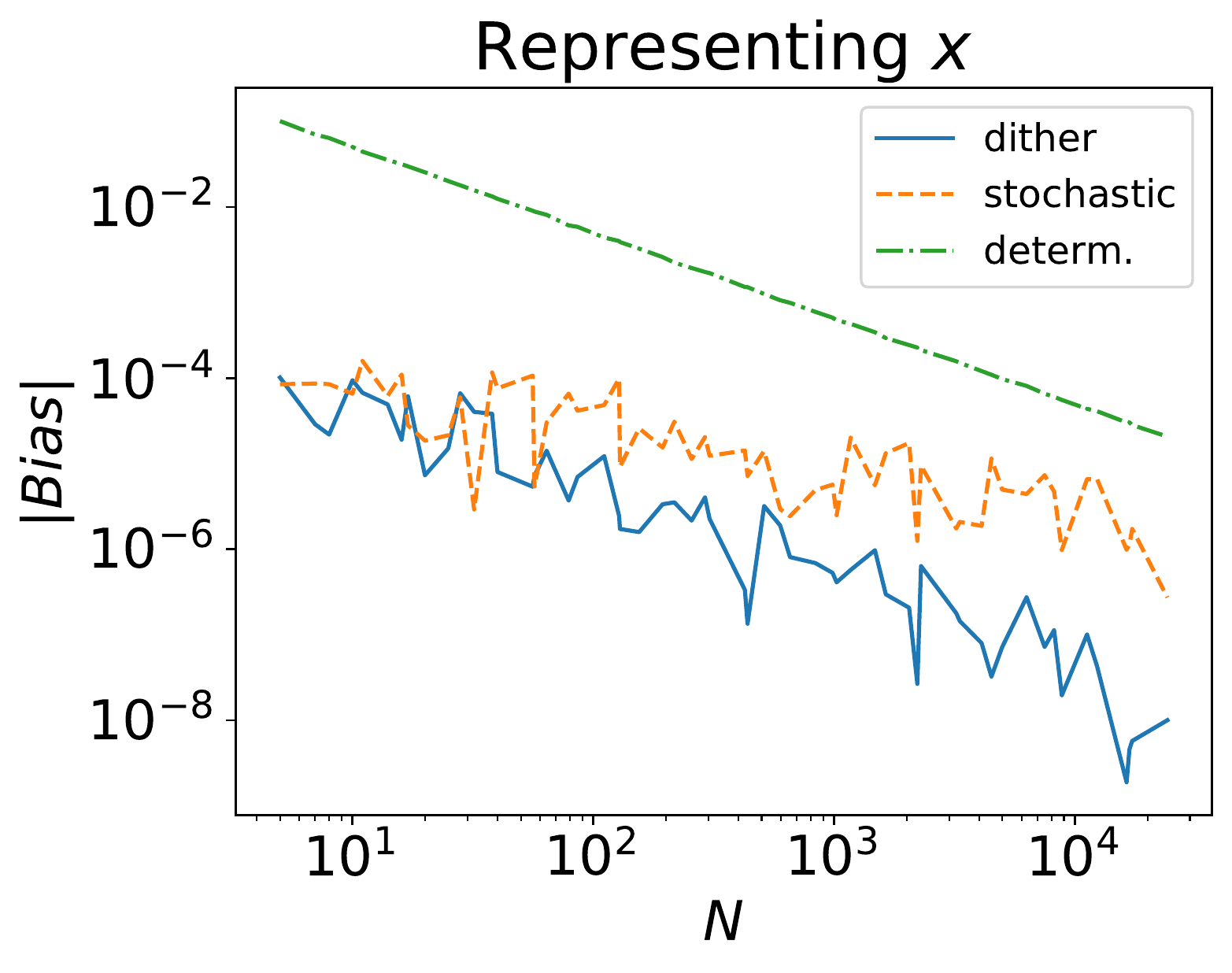}}
\caption{Sample estimate of $|Bias|$ to represent $x$ for various values of $N$.}
\label{fig:two}
\end{figure}

\begin{figure}[htbp]
\centerline{\includegraphics[width=2.71in]{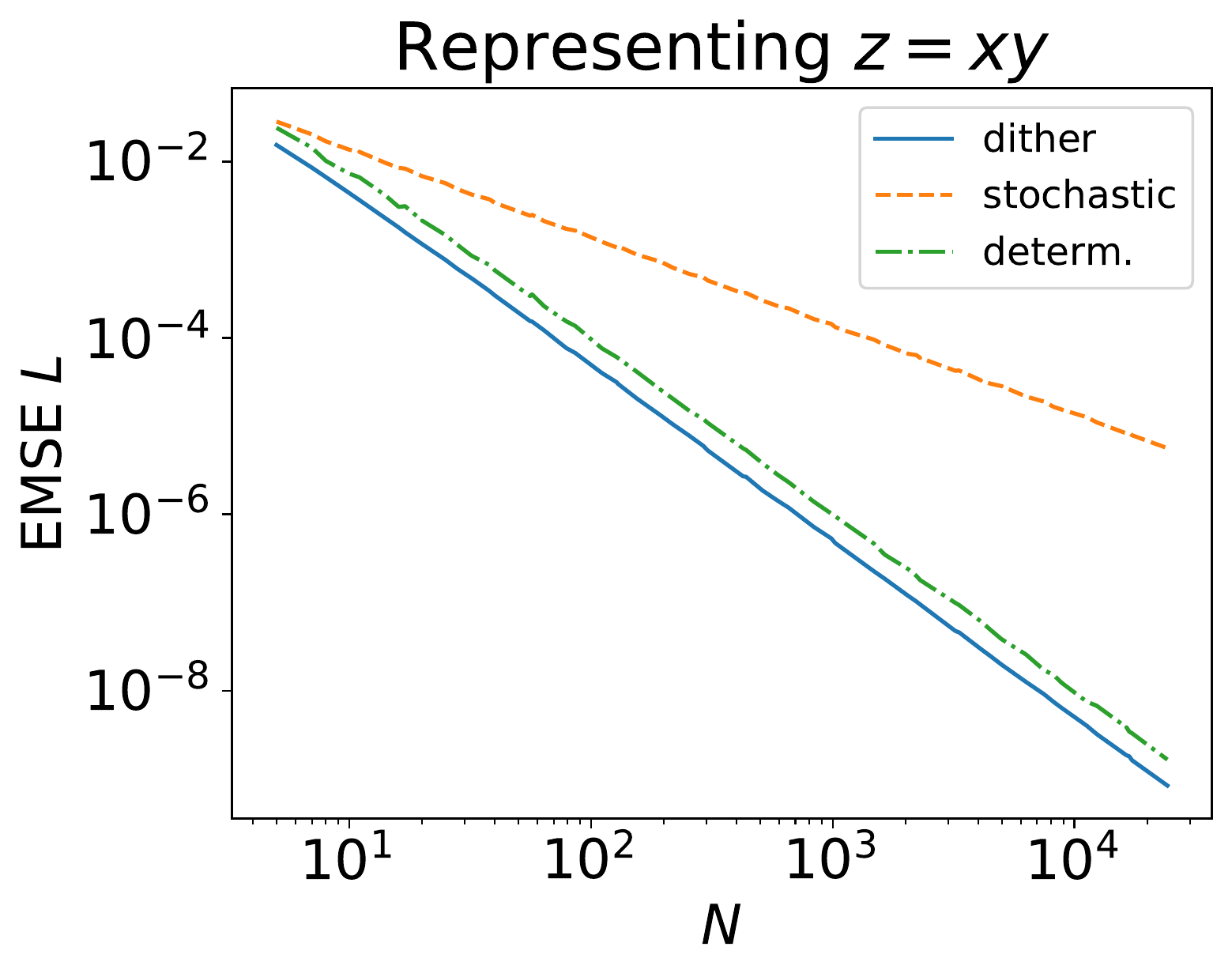}}
\caption{Sample estimate of EMSE $L$ to represent $z=xy$ for various values of $N$.}
\label{fig:three}
\end{figure}

\begin{figure}[htbp]
\centerline{\includegraphics[width=2.71in]{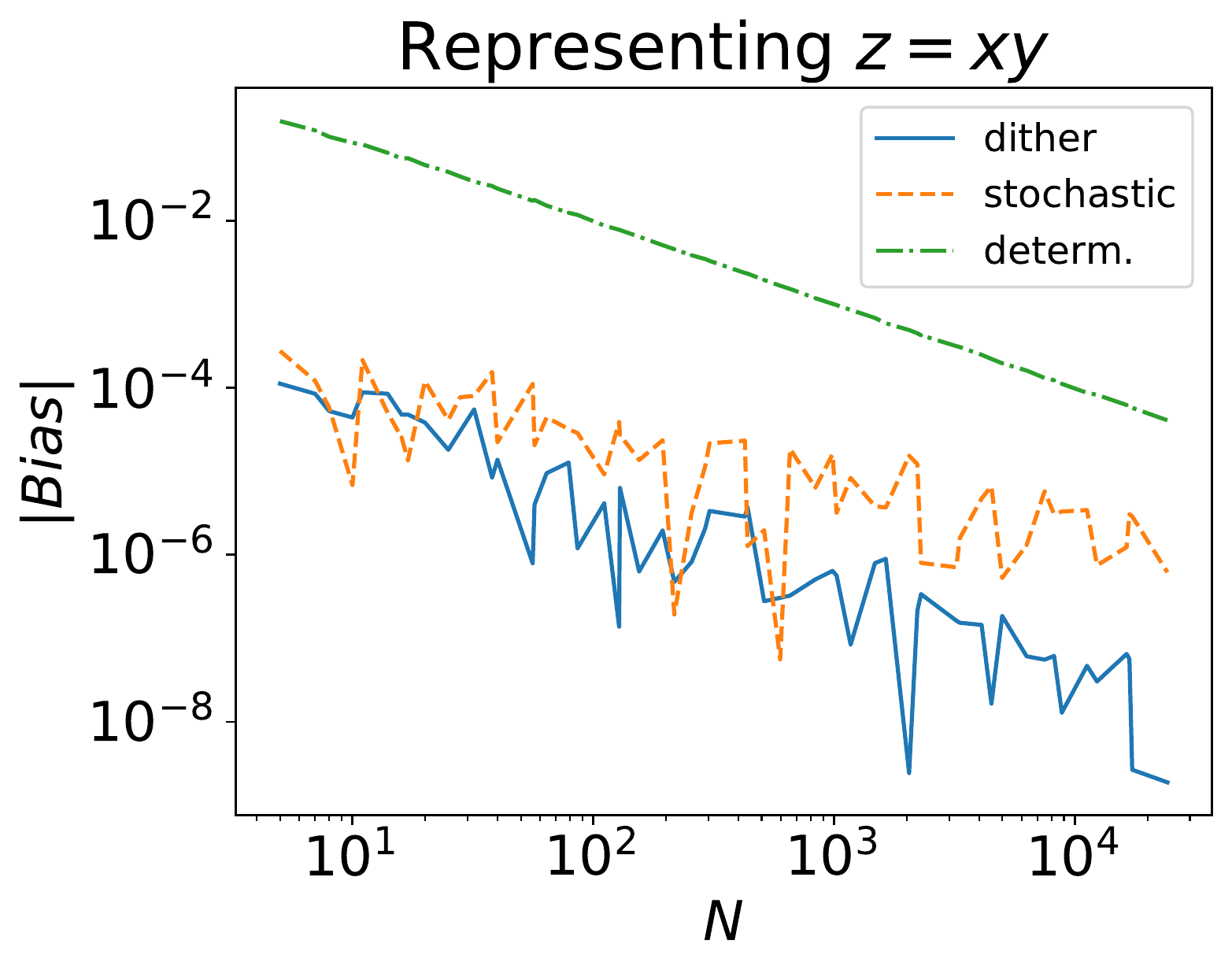}}
\caption{Sample estimate of $|Bias|$ to represent $z=xy$ for various values of $N$.}
\label{fig:four}
\end{figure}

\begin{figure}[htbp]
\centerline{\includegraphics[width=2.71in]{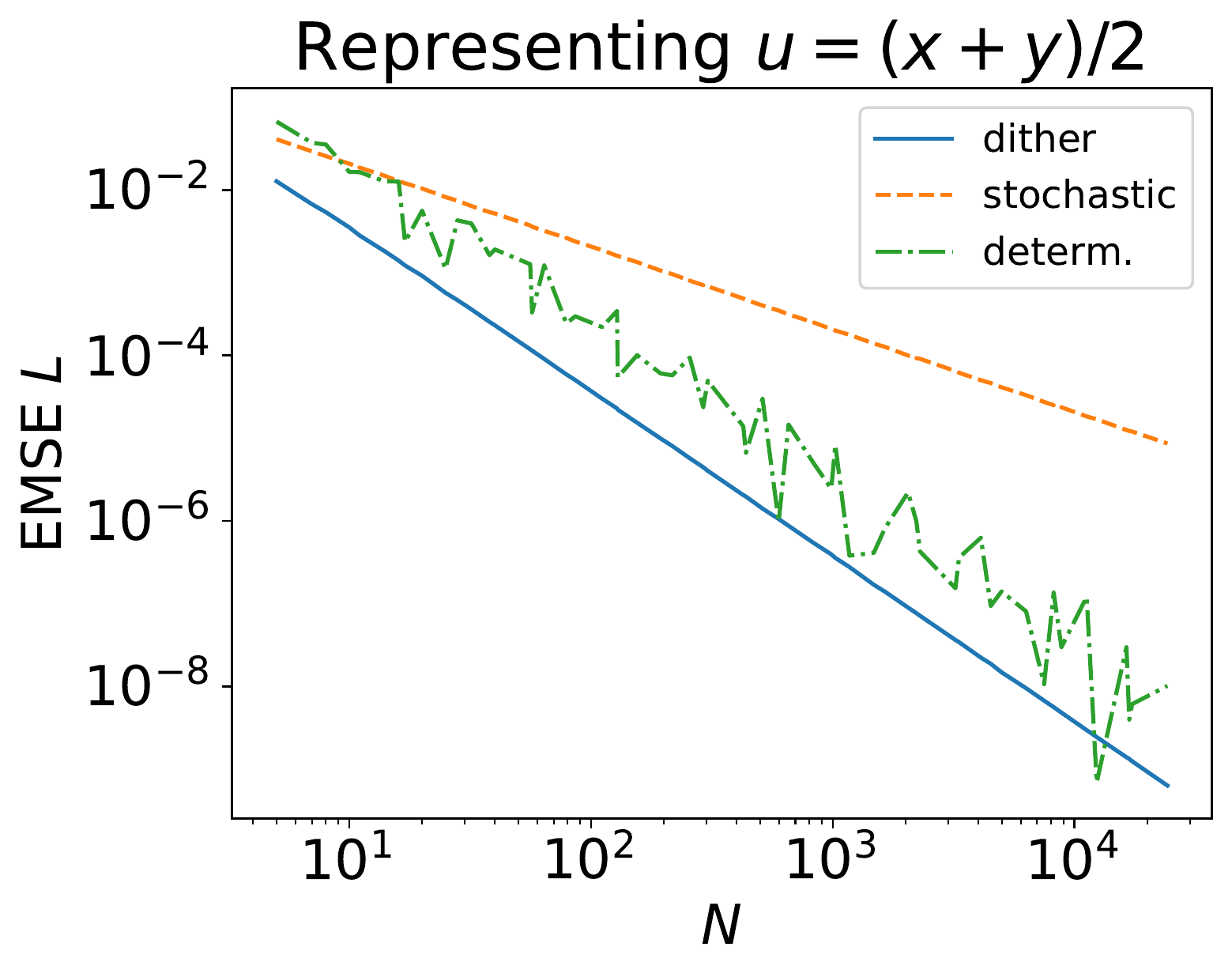}}
\caption{Sample estimate of EMSE $L$ to represent $u=\frac{x+y}{2}$ for various values of $N$.}
\label{fig:five}
\end{figure}

\begin{figure}[htbp]
\centerline{\includegraphics[width=2.71in]{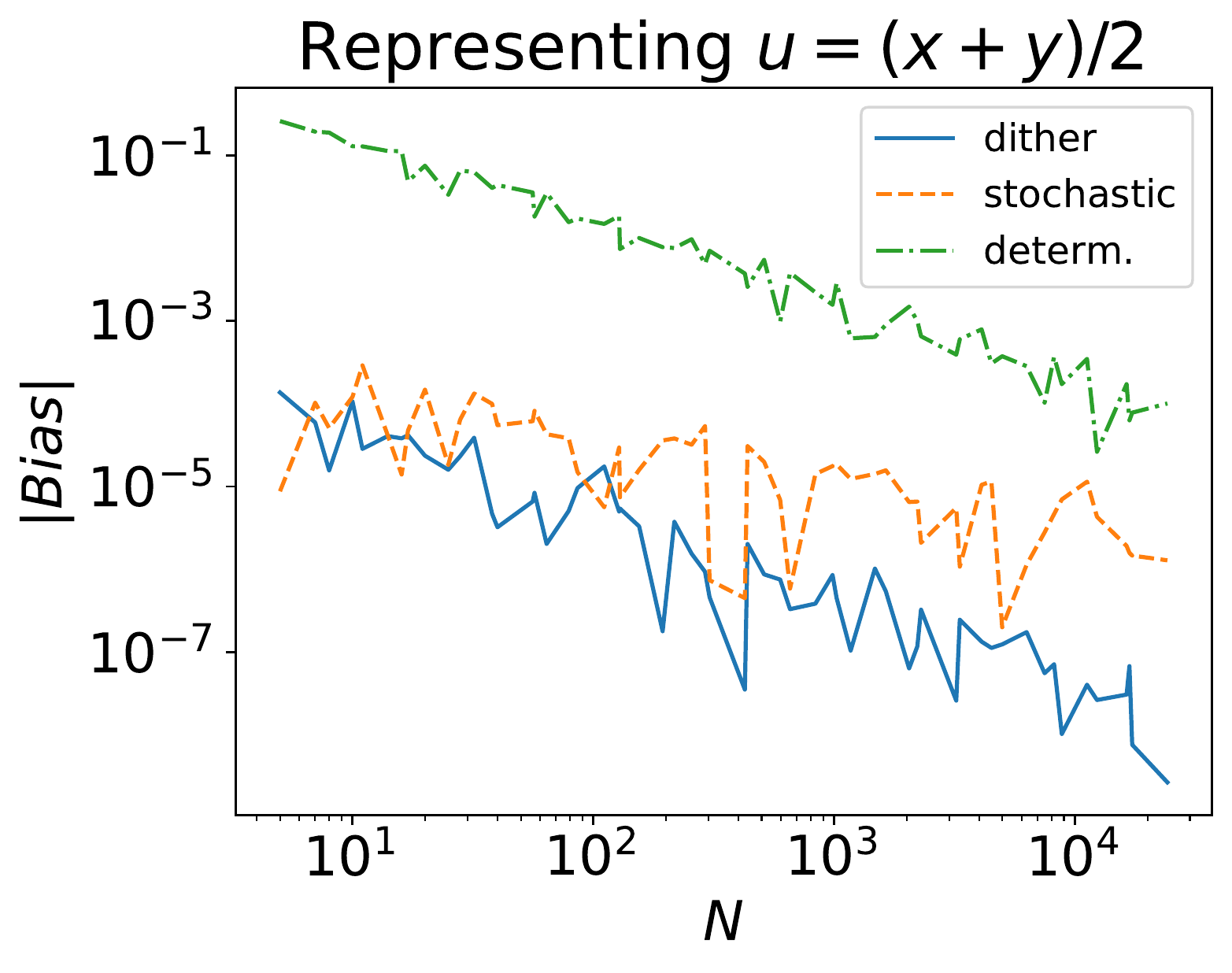}}
\caption{Sample estimate of $|Bias|$ to represent $u=\frac{x+y}{2}$ for various values of $N$.}
\label{fig:six}
\end{figure}

In Figs. \ref{fig:one}-\ref{fig:six} we show the EMSE $L$ and the bias for the computing schemes above by generating $1000$ independent pairs $(x,y)$ from a uniform distribution of $X$ and of $Y$. For each pair $(x,y)$, 1000 trials of dither computing and stochastic computing\footnote{The set of pairs $(x,y)$ are the same for the 3 schemes. For the deterministic variant, only 1 trial is performed as $X$ are $Y$ are deterministic.} are used to represent them and compute the product $z$ and average $u$.

We see that the sample estimate for the bias for $x$, $z$ and $u$ are lower for both the stochastic computing scheme and the dither computing scheme as compared with the deterministic variant. On the other hand, the dither computing scheme has similar EMSE on the order of $O(\frac{1}{N^2})$ as the deterministic scheme, whereas the stochastic computing scheme has higher EMSE on the order of $\Omega(\frac{1}{N})$.

Even though both stochastic computing and dither computing have zero bias, the sample estimate of this bias is lower for dither computing than for stochastic computing. This is because the standard error of the mean (SEM) is proportional to the standard deviation and dither computing has standard deviation $O(\frac{1}{N})$ vs $\Omega(\frac{1}{\sqrt{N}})$ for stochastic computing and this is observed in the slope of the curves in Figs \ref{fig:two}, \ref{fig:four}, \ref{fig:six}. 

Furthermore, even though the dither computing representations of $x$ and $y$ have worse EMSE than the deterministic variant, the dither computing representations of both the product $z$ and the scaled addition $u$ have better EMSE.
The asymptotic behavior of bias and EMSE for these different schemes are listed in Table \ref{tbl:one}.

\begin{table}
\begin{center}
\begin{tabular} {|c||c|c|c|}
\hline
& Stoch. Comp. & Determ. Variant & Dither Comp. \\ [0.5ex]
\hline\hline
Bias (repr.) & $0$ & $\Theta(\frac{1}{N})$ &  $0$ \\
\hline
Variance (repr.)  & $\Omega(\frac{1}{N})$ & $0$ & $\Theta(\frac{1}{N^2})$ \\
\hline
EMSE $L$ (repr.)   & $\Omega(\frac{1}{N})$ & $\Theta(\frac{1}{N^2})$ & $\Theta(\frac{1}{N^2})$ \\
\hline\hline
Bias (mult.) & $0$ & $\Theta(\frac{1}{N})$ & $0$ \\
\hline
Variance (mult.)  & $\Omega(\frac{1}{N})$ & $0$ & $\Theta(\frac{1}{N^2})$ \\
\hline
EMSE $L$ (mult.)   & $\Omega(\frac{1}{N})$ & $\Theta(\frac{1}{N^2})$ & $\Theta(\frac{1}{N^2})$ \\
\hline\hline
Bias (average) & $0$ & $\Theta(\frac{1}{N})$ & $0$ \\
\hline
Variance (average)  & $\Omega(\frac{1}{N})$ & $0$ & $\Theta(\frac{1}{N^2})$ \\
\hline
EMSE $L$ (average)  & $\Omega(\frac{1}{N})$ & $\Theta(\frac{1}{N^{2}})$ & $\Theta(\frac{1}{N^2})$ \\
\hline
\end{tabular}
\end{center}
\caption{Asymptotic behavior of Bias, Variance and EMSE for stochastic computing, deterministic variant and dither computing to represent a number, to multiply 2 numbers and to perform the average (scaled addition) operation.}\label{tbl:one}
\end{table}

\section{Asymmetry in operands}
In the dither computing scheme (and in the deterministic variant of the stochastic computing as well), the encoding of the two operands $x$ and $y$ are different. For instance $x$ is encoded as a unary number (denoted as {\em Format 1}) and $y$ has its $1$-bits spread out as much as possible (denoted as {\em Format 2}) for multiplication while both $x$ and $y$ are encoded as unary numbers for scaled addition. For multilevel arithmetic operations, this asymmetry requires additional logic to convert the output of multiplication and scaled addition into these 2 formats depending on which operand and which operation the next arithmetical operation is.
On the other hand, there are several applications where the need for this additional step is reduced. For instance,

\begin{enumerate}
\item 
In memristive crossbar arrays \cite{Gokmen2017}, the sequence of pulses in the product is integrated and converted to digital via an A/D converter and thus the product sequence of pulses is not used in subsequent computations.
\item 
Using stochastic computing to implement the matrix-vector multiply-and-add in neural networks \cite{liu:2020}, one of the operand is always a weight or a bias and thus fixed throughout the inference operation. Thus the weight can be precoded in Format 2 for multiplication and the bias value is precoded in Format 1 for addition, whereas the data to be operated on is always in Format 1 and the result recoded to Format 1 for the next operation.
\end{enumerate} 

\section{Dither rounding: stochastic rounding revisited}\label{sec:dround}
In order to reduce power while increasing throughput, there has been much research activities in using reduced precision hardware, in particular in deep learning both for training and inference \cite{Colangelo2018,Choi2019}. In these applications, the data and operations are performed with far lower precision than traditional fixed point or floating point arithmetic, going to as low as a single bit \cite{Qin2020}. To address the loss of precision in such reduced precision arithmetical units, stochastic rounding has emerged as an alternative mechanism to deterministic rounding for using reduced precision hardware in applications such as solving differential equations \cite{hopkins:2020} and deep learning \cite{Gupta2015a}.
As mentioned in Sec. \ref{sec:sround}, 1-bit stochastic rounding can be considered as the special case of stochastic computing with $N=1$.
For $k$-bit stochastic rounding, the situation is similar as only the least significant bit is stochastic. Another alternative interpretation is that stochastic computing is stochastic rounding in time, i.e. $X_i$, $i=1,\cdots, N$ can be considered as applying stochastic rounding $N$ times. Since the standard error of the mean of dither computing is asymptotically superior to 
stochastic computing, we expect this advantage to persist for rounding as well when applied over time. 

Thus we introduce {\em dither rounding} as follows. We assume $\alpha\geq 0$ as the case $\alpha <0$ can be handled similarly. We define
dither rounding of a real number $\alpha \geq 0$ as
$d(\alpha,i) = \lfloor \alpha\rfloor + X_i$ where $\{X_i\}$ is the dither computing representation of $x = \alpha-\lfloor \alpha\rfloor$ as defined in Sect. \ref{sec:dither} and $\alpha-\lfloor \alpha\rfloor$ is the fractional part of $\alpha$.
Note that there is an index $i$ in the definition of $d(\cdot, \cdot)$ which is an integer $0\leq i < N$. In practice we will compute $i$ as $\sigma(i_s \mod N)$, where $i_s$ counts how many times the dither rounding operation has been applied so far and $\sigma$ is a fixed permutation, one for the left operand and one for the right operand of the scalar multiplier. The implementation complexity of dither rounding is similar to stochastic rounding, except that we need to keep track of the index $i$, whereas in stochastic rounding, the rounding of the elements are done independently.

\begin{figure}[htbp]
\centerline{\includegraphics[width=3.66in]{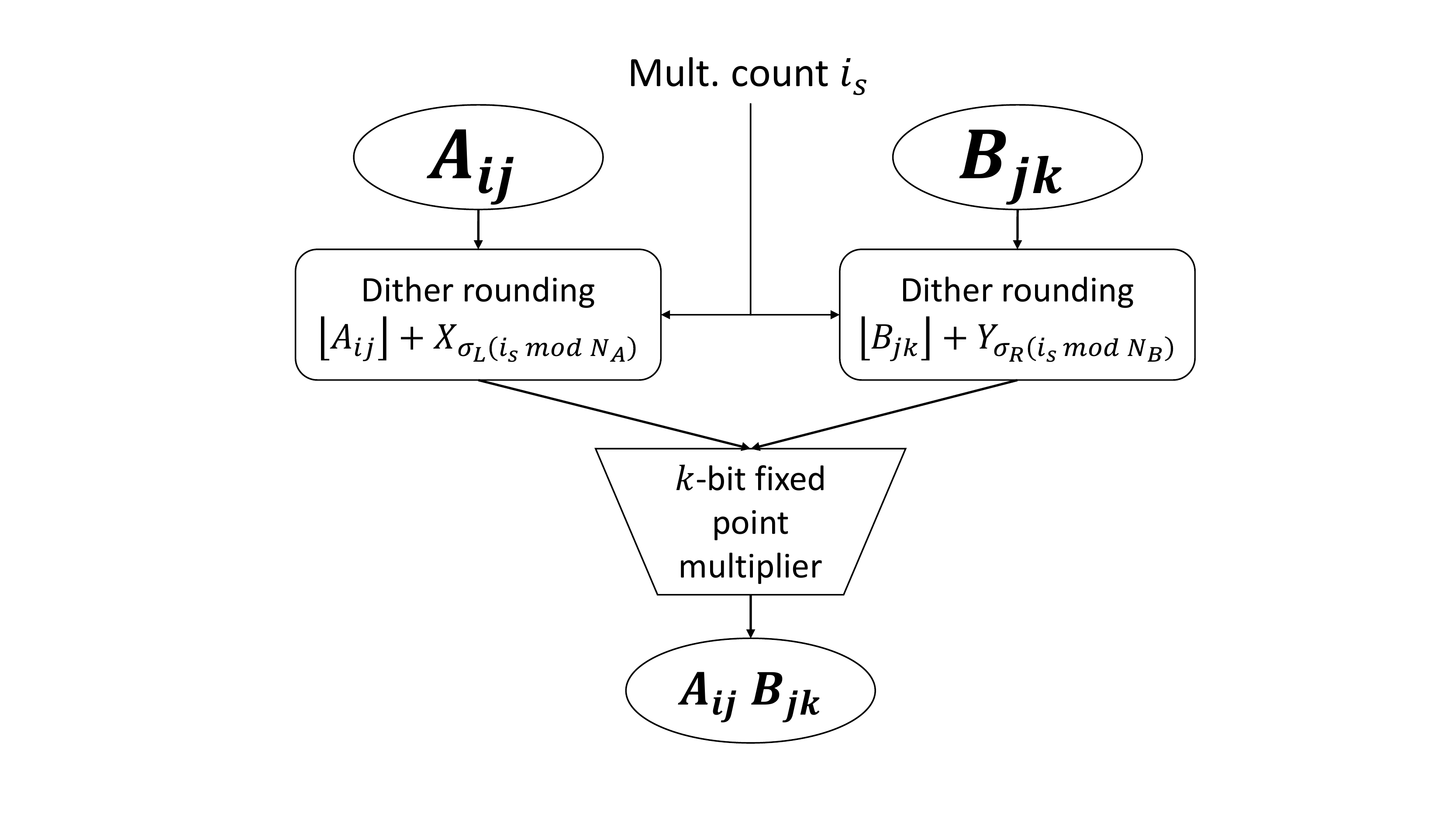}}
\caption{Dither rounding to compute the partial result $A_{ij}B_{jk}$. $X_u$ and $Y_v$ are the dither computing representations of $A_{ij}-\lfloor A_{ij}\rfloor$ and $B_{jk}-\lfloor B_{jk}\rfloor$.}
\label{fig:dround-system}
\end{figure}

To illustrate the performance of these different rounding schemes, consider the problem of matrix-matrix multiplication, a workhorse of computational science and deep learning algorithms.
Let $A$ and $B$ be $p\times q$ and $q\times r$ matrices with elements in $[0,1]$. We denote the $(i,j)$-th element of $A$ as $A_{ij}$.
The goal is to compute the matrix $C = AB$.  
A straightforward algorithm for computing $C$ requires $pqr$ (scalar) multiplications. Although there are asympotically more optimal algorithms for square matrices such as Strassen \cite{strassen:1969}, Coppersmith-Winograd \cite{Coppersmith1990} and Alman-Williams \cite{Alman2021} they are not widely used in practice. Let us assume that we have at our disposal only $k$-bit fixed point digital multipliers and thus floating point real numbers are rounded to $k$-bits before using the multiplier. We use
the following standard definition of a $k$-bit quantizer. For simplicity, we will only deal with nonnegative numbers. The quantized value is simply $q(x) = round(x)$ for $x\in [0,2^k-1]$.
If $x < 0$, then $q(x) = 0$ (underflow) and if $x > 2^k-1$, then $q(x) = 2^k-1$ (overflow).  We will also refer to this as traditional rounding. We want to compare the performance of computing $C=AB$ between traditional rounding, stochastic rounding and dither rounding. In particular, since each element of $A$ is used $r$ times and each element of $B$ is used $p$ times, for dither rounding we set $N = N_A = r$ for matrix $A$ and $N = N_B = p$ for matrix $B$. For dither rounding the computation of each of the $pqr$ partial results $A_{ij}B_{jk}$ is illustrated in Fig. \ref{fig:dround-system}, and the other schemes can be obtained by simply replacing the rounding scheme.
We measure the error by computing the Frobenius matrix norm $e_f = \|C-\tilde{C}\|_{F}$ where $\tilde{C}$ is the product matrix computed using the specified rounding method and the $k$-bit fixed point multiplier. In our case this is implemented by rescaling the interval $[0,1]$ to $[0,2^k-1]$ and rounding to fixed point $k$-bit integers. Note that the Frobenius matrix norm is equivalent to the $l^2$ vector norm when the matrix is flattened as a vector.

In Sect. \ref{sec:sround}, it was shown that deterministic rounding has the lowest EMSE, but Ref. \cite{connolly:2020} argues that correlation in subsequent data makes an unbiased estimator such as stochastic rounding a better choice for deep learning. We propose another reason why dither or stochastic rounding can be superior to deterministic rounding in deep learning. If the prior distribution of the data is not uniform, the output levels of the quantizer are not obtained uniformly and some of the output levels may be wasted by rarely being used. As an extreme example, if the input is in the range $[0,\frac{1}{2})$, the output of a $k$-bit quantizer in deterministic rounding will always be $0$ and all information about the input is lost. On the other hand, the output from a dither or stochastic rounding will take on both values $0$ and $1$.
  
If we assume that the range of the matrix elements is narrow when compared to the quantization interval, then we expect dither rounding (and stochastic rounding) to outperform traditional rounding.
For example, take the special case of $A = \alpha J$ and $B = \beta J$, where $J$ is the square matrix of all $1$'s and $\alpha, \beta \in [0,1]$. 
When we use traditional rounding to round the elements of $A$ and $B$, the corresponding $\tilde{C}$ is $\gamma J^2$, where $\gamma = round((2^k-1) \alpha)\cdot round((2^k-1)\beta)/(2^k-1)^2$ which is $\neq \alpha\beta$ in general.
The analysis in Section \ref{sec:mult} shows that for both dither rounding and stochastic rounding the resulting $\tilde{C}$ satisfies $E(\tilde{C}) = \alpha\beta J^2 = AB$, with $E(e_f) = \Theta(\frac{1}{N})$ for dither rounding and $E(e_f) = \Theta(\frac{1}{\sqrt{N}})$ for stochastic rounding.

\begin{figure}[htbp]
\centerline{\includegraphics[width=2.71in]{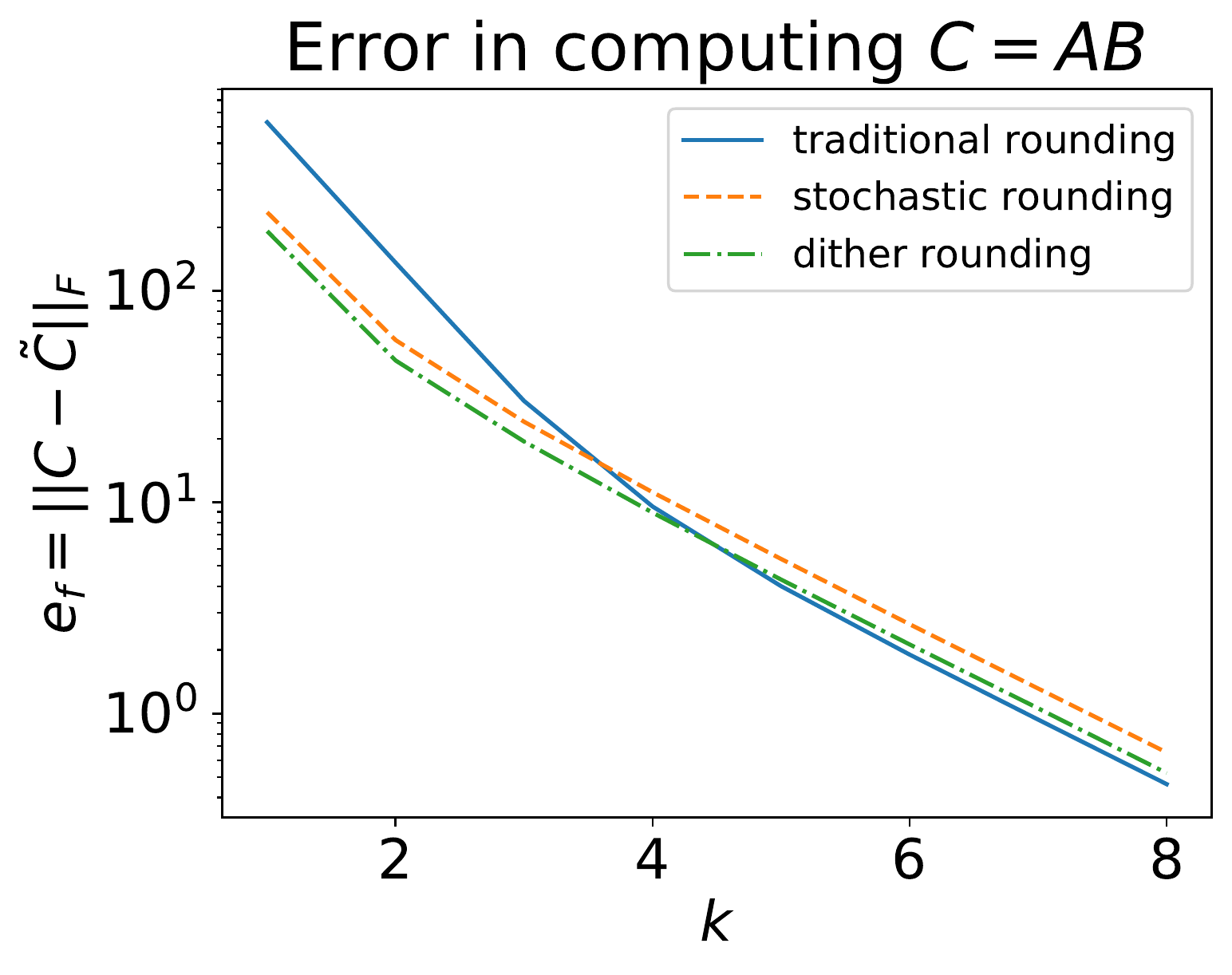}}
\caption{Comparison of various rounding methods for multiplying two $100$ by $100$ matrices with entries in $[0,0.5)$.}
\label{fig:dround-lv}
\end{figure}

To ensure there is no overflow or underflow in the quantization process, in practical applications such as deep learning training the numbers are conservatively scaled to lie well within the range of the quantizer so the above assumption is reasonable. As an another example, we generate 100 pairs of 100 by 100 matrices $A$ and $B$ where elements of $A$ and $B$ are randomly chosen from the range $[0,\frac{1}{2})$ and choose $N=100$. The average $e_f$ for traditional rounding, stochastic computing and dither computing are shown in Fig. \ref{fig:dround-lv}.\footnote{Note that for traditional rounding and $k=1$, $A$ and $B$ are both rounded to the zero matrix, and $e_f = \|AB\|_{F}$ in this case.} We see that dither rounding has smaller $e_f$ than stochastic rounding and that for small $k$ both dither computing and stochastic rounding has  significant lower error in computing $AB$ than traditional rounding. There is a threshold $\tilde{k}$ where traditional rounding outperforms dither or stochastic rounding for $k\geq \tilde{k}$, and we expect this threshold to increase when $N,p,q,r$ increase.

\begin{figure}[htbp]
\centerline{\includegraphics[width=2.71in]{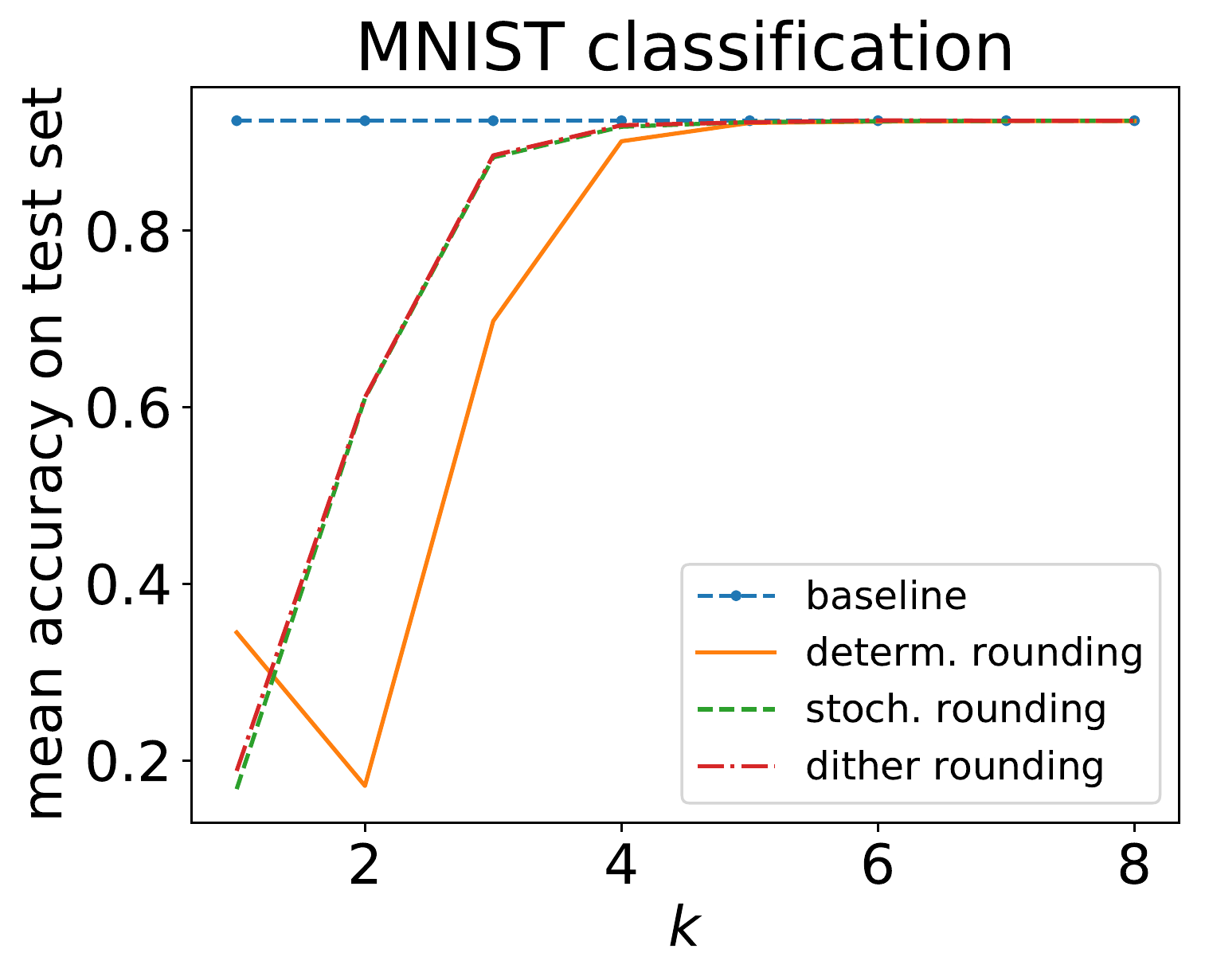}}
\caption{Comparison of deterministic, dither and stochastic rounding for the MNIST handwritten digits classification task. The mean accuracy is computed over $1000$ trials.}
\label{fig:mnist}
\end{figure}

\begin{figure}[htbp]
\centerline{\includegraphics[width=2.71in]{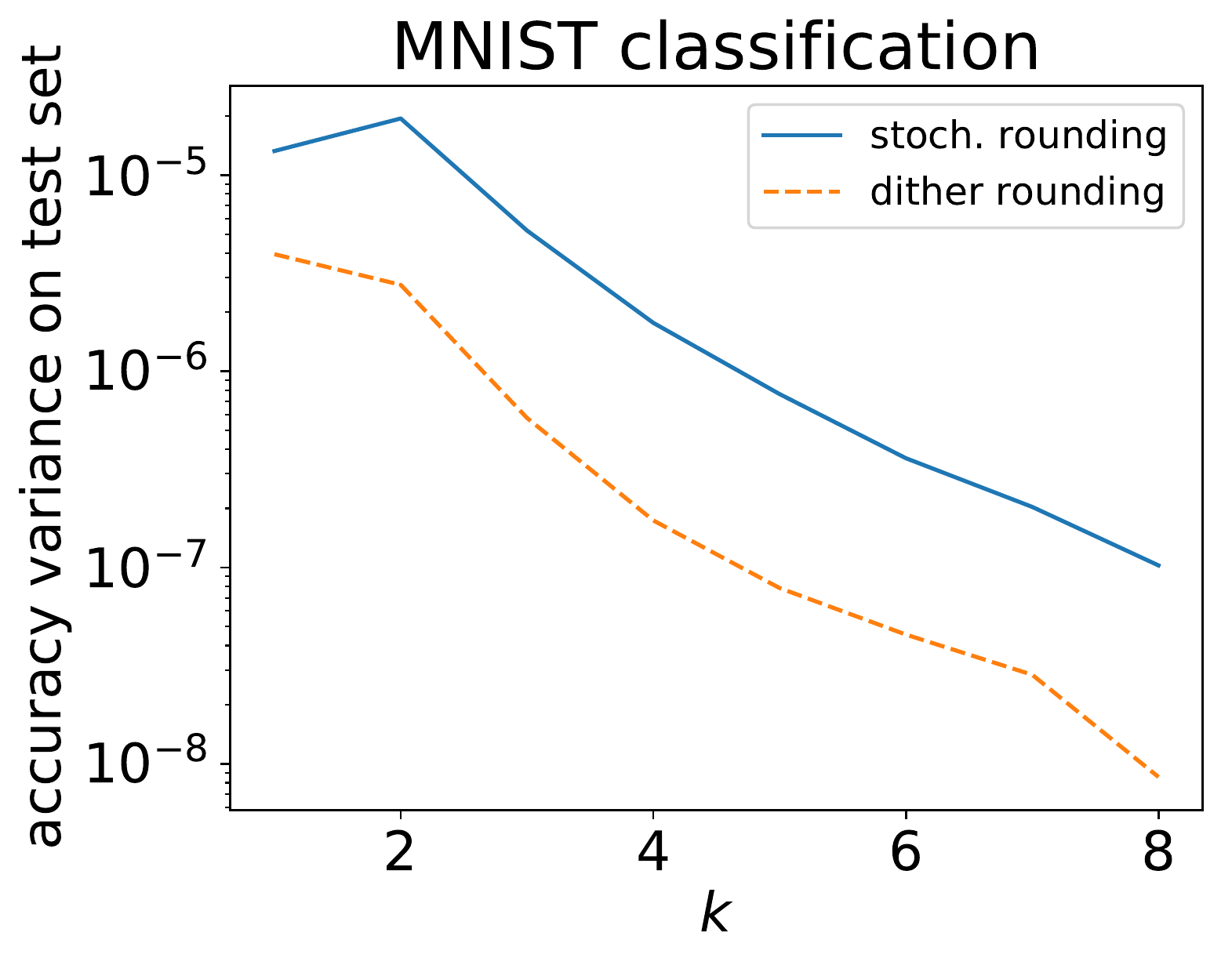}}
\caption{Comparison of the variance in MNIST classification accuracy for dither and stochastic rounding. The sample variance is computed over $1000$ trials.}
\label{fig:mnistvar}
\end{figure}

Next, we apply this approach to a simple neural network solving the MNIST handwritten digits recognition task \cite{LeCun2010}. 
The input images are $28\times 28$ grayscale images with pixel values in the range $[0, 1]$.
A single layer neural network with a softmax function can obtain an accuracy of $92.4\%$ on the $10000$ sample test set, which we denote as the baseline accuracy. The neural network has a single weight matrix and a single bias vector and inference includes a single matrix-matrix multiplication of the input data matrix and the weight matrix.
We scaled the weight matrix to the range $[-1,1]$. In order to use a $k$-bit fixed point multiplier, we rescale both the weights and the input from $[-1,1]$ to $[0,2^k-1]$ and apply the standard $k$-bit quantizer. Note that since the input is restricted to the range $[0,1]$, it did not fully utilize the full range of the quantizer. 
We apply the 3 rounding methods discussed earlier to compute the partial products in the matrix multiplication (Fig. \ref{fig:mnist}). We see in Fig. \ref{fig:mnist} that dither rounding has similar classification accuracy as stochastic rounding and both of them are significantly better than deterministic rounding for small $k > 1$. Note that the rounding method sometimes has slightly better accuracy than using full precision.
Furthermore, dither rounding has less variance on the classification accuracy compared with stochastic rounding (Fig. \ref{fig:mnistvar}).

\section{Other dither rounding schemes for matrix multiplication}
\begin{figure}[htbp]
\centerline{\includegraphics[width=2.71in]{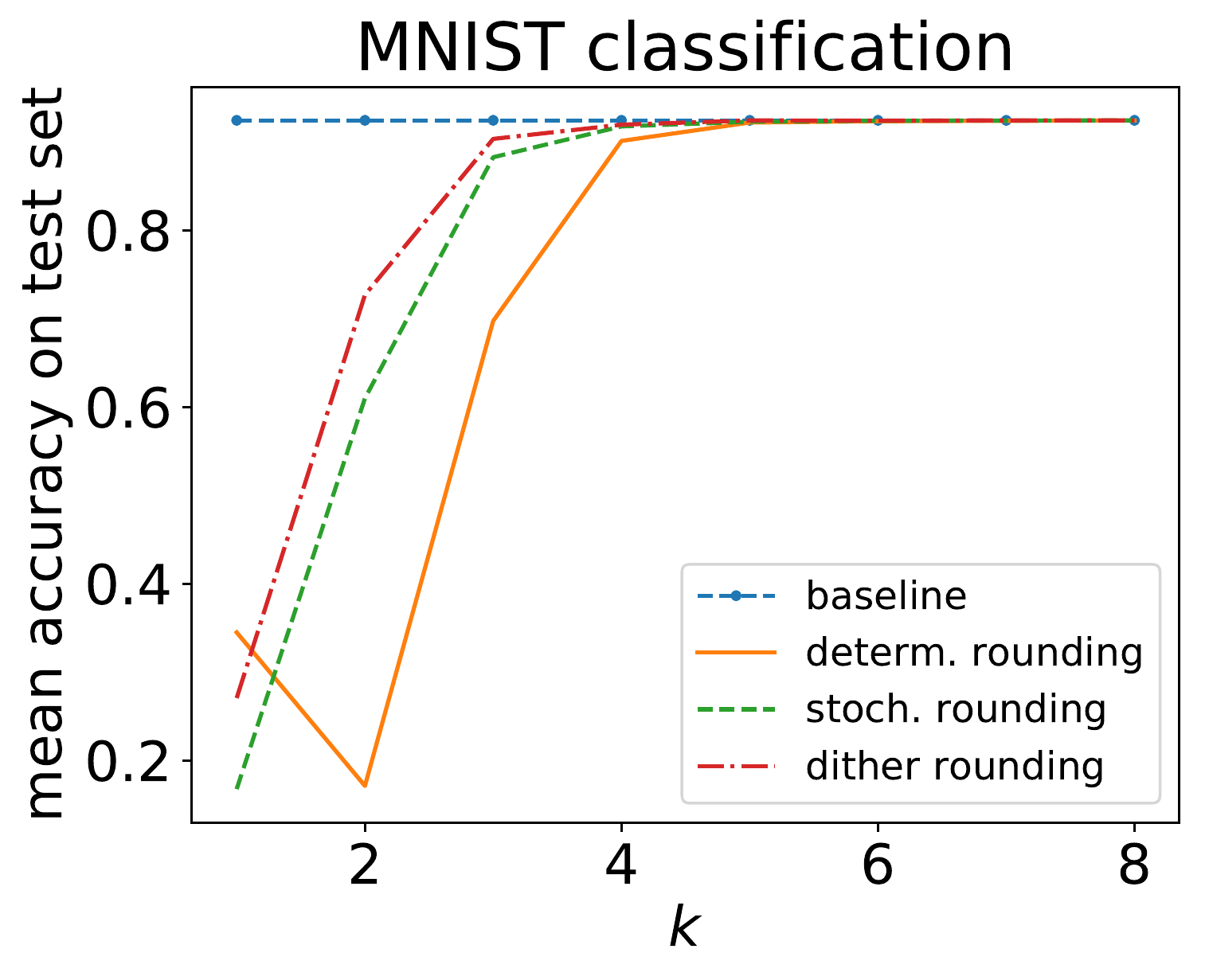}}
\caption{Comparison of rounding schemes for the MNIST classification task. The dither rounding is performed using a variant where the input are rounded once. The mean accuracy is computed over $1000$ trials.}
\label{fig:mnistvariant}
\end{figure}

\begin{figure}[htbp]
\centerline{\includegraphics[width=2.71in]{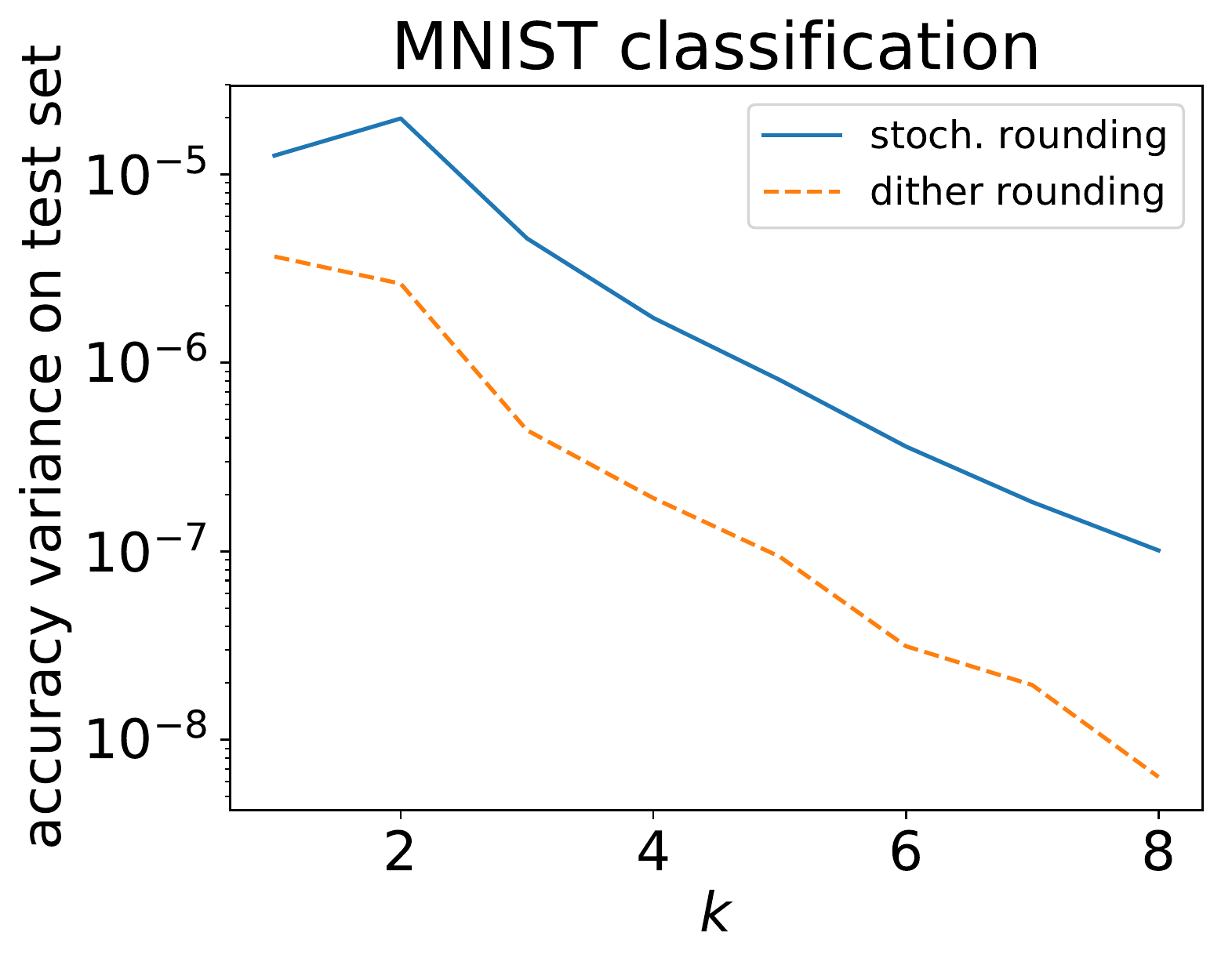}}
\caption{Classification accuracy variance for MNIST using a dither rounding variant where the input are rounded once. The sample variance is computed over $1000$ trials.}
\label{fig:mnistvariantvar}
\end{figure}

In the above matrix multiplication scheme, the dither (or stochastic) rounding operation is performed on each of the $pqr$ partial products and 2 rounding operations per partial product (Fig. \ref{fig:dround-system}), resulting in $2pqr$ rounding operations.
We next consider other variants for dither rounding schemes for matrix multiplication. For instance, one can compute the partial product $A_{ij}B_{jk}$ where $A_{ij}$ is rounded once for each $(i,j)$ and applied to each $k$ whereas $B_{jk}$ is rounded for each partial product. For the MNIST task this corresponds to the input being only quantized once. We find that this dither rounding variant results in slighter better performance than stochastic rounding as shown in Figs. \ref{fig:mnistvariant} and \ref{fig:mnistvariantvar}. The number of rounding  operations is $pq + pqr = pq(r+1)$.

\begin{figure}[htbp]
\centerline{\includegraphics[width=3in]{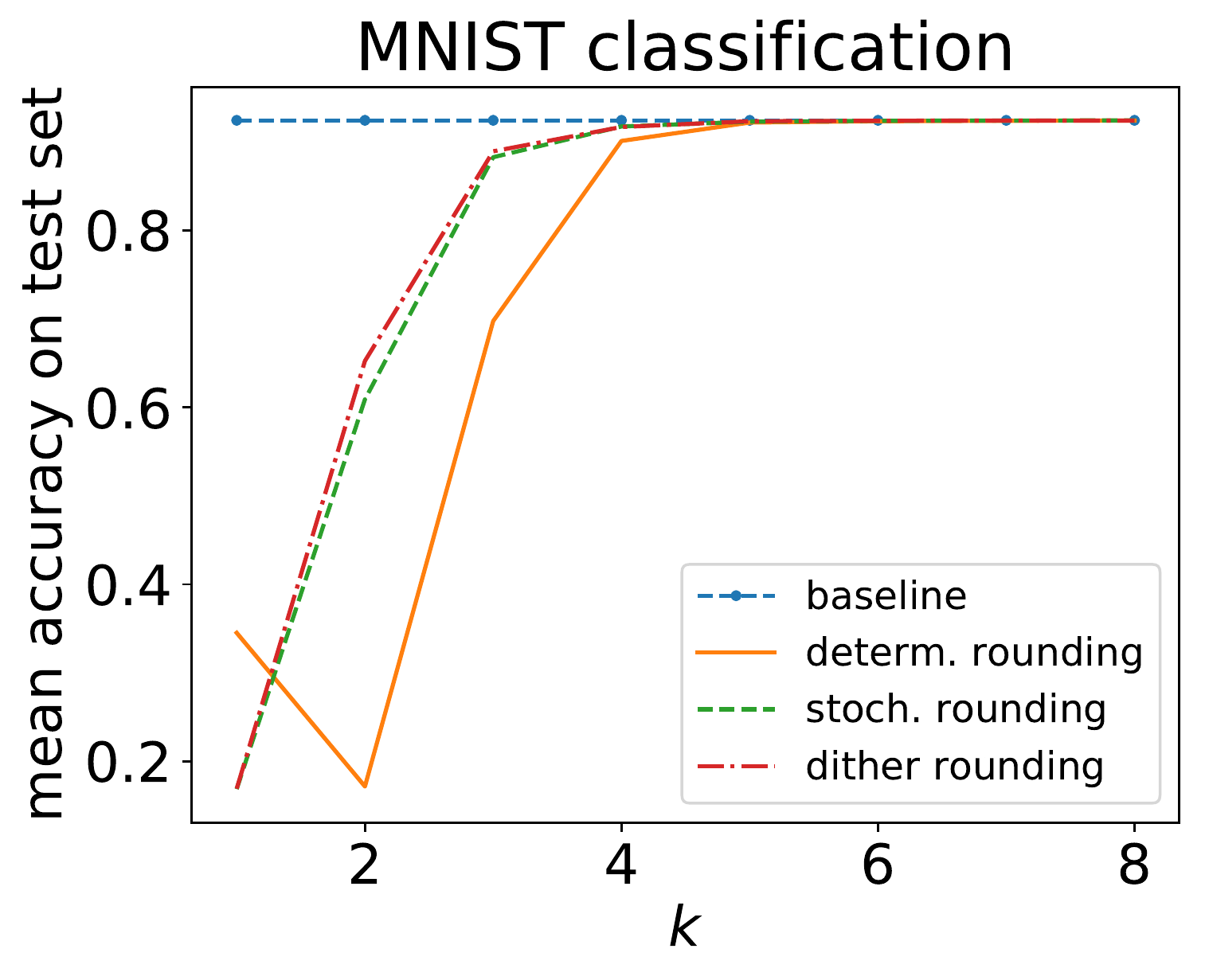}}
\caption{Comparison of deterministic, dither and stochastic rounding for MNIST by quantizing the matrices separately. The mean accuracy is computed over $1000$ trials.}
\label{fig:mnistseperate}
\end{figure}

\begin{figure}[htbp]
\centerline{\includegraphics[width=3in]{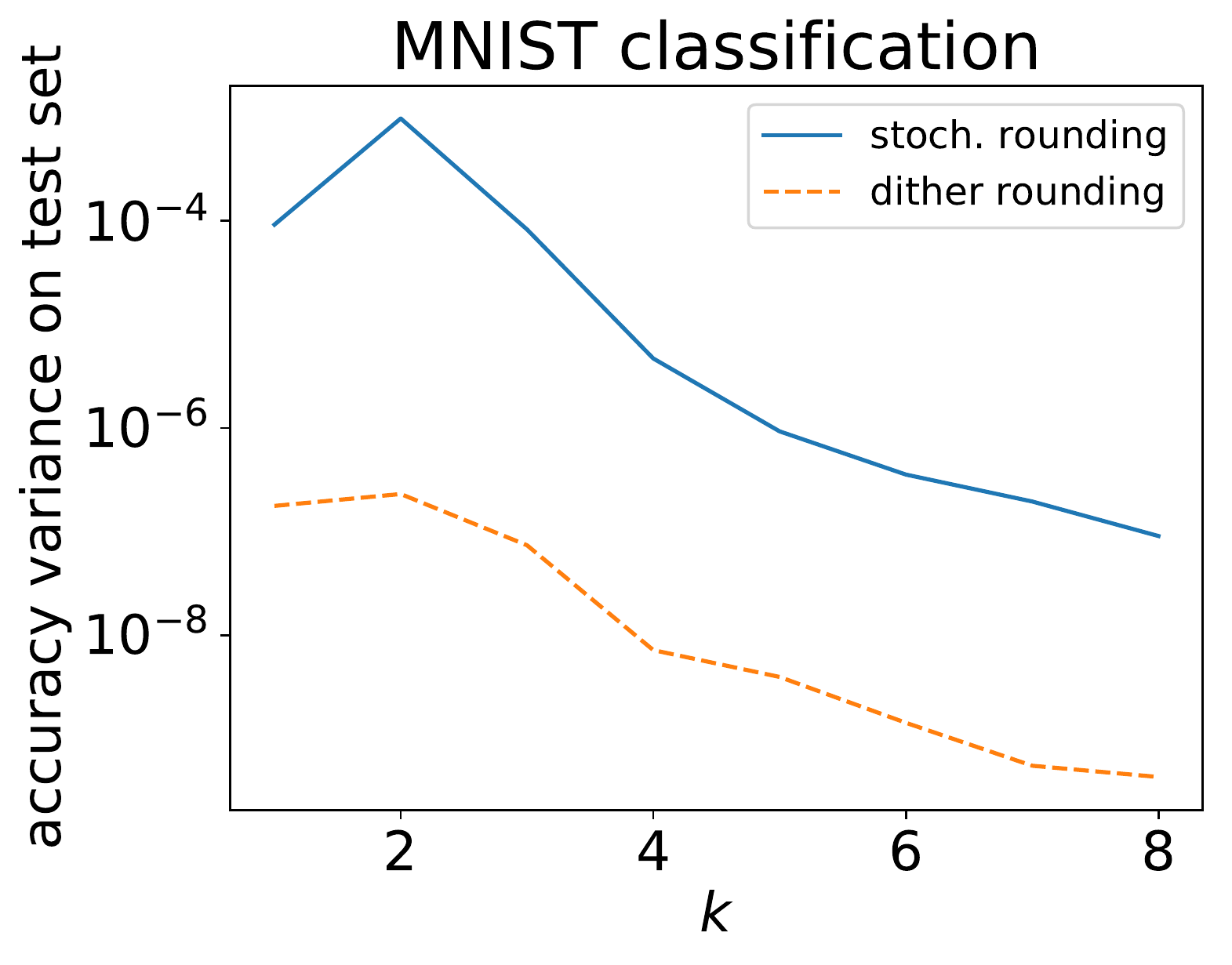}}
\caption{Classification accuracy variance in MNIST classification by quantizing the matrices separately. The sample variance is computed over $1000$ trials.}
\label{fig:mnistseperatevar}
\end{figure}

Another variant is to apply the rounding to $A$ and $B$ separately and then perform matrix multiplication on the rounded matrices. This will only require $pq+qr= (p+r)q$ rounding operations which can be much less that $2pqr$. We show that this variant (for both dither and stochastic rounding) can still provide benefits when compared with deterministic rounding.
The results are shown in Figs. \ref{fig:mnistseperate}-\ref{fig:mnistseperatevar}. For stochastic and dither rounding, the mean accuracy over $1000$ trials are plotted. 
We see that the results are similar to Figs. \ref{fig:mnist}-\ref{fig:mnistvar}.

\begin{figure}[htbp]
\centerline{\includegraphics[width=3in]{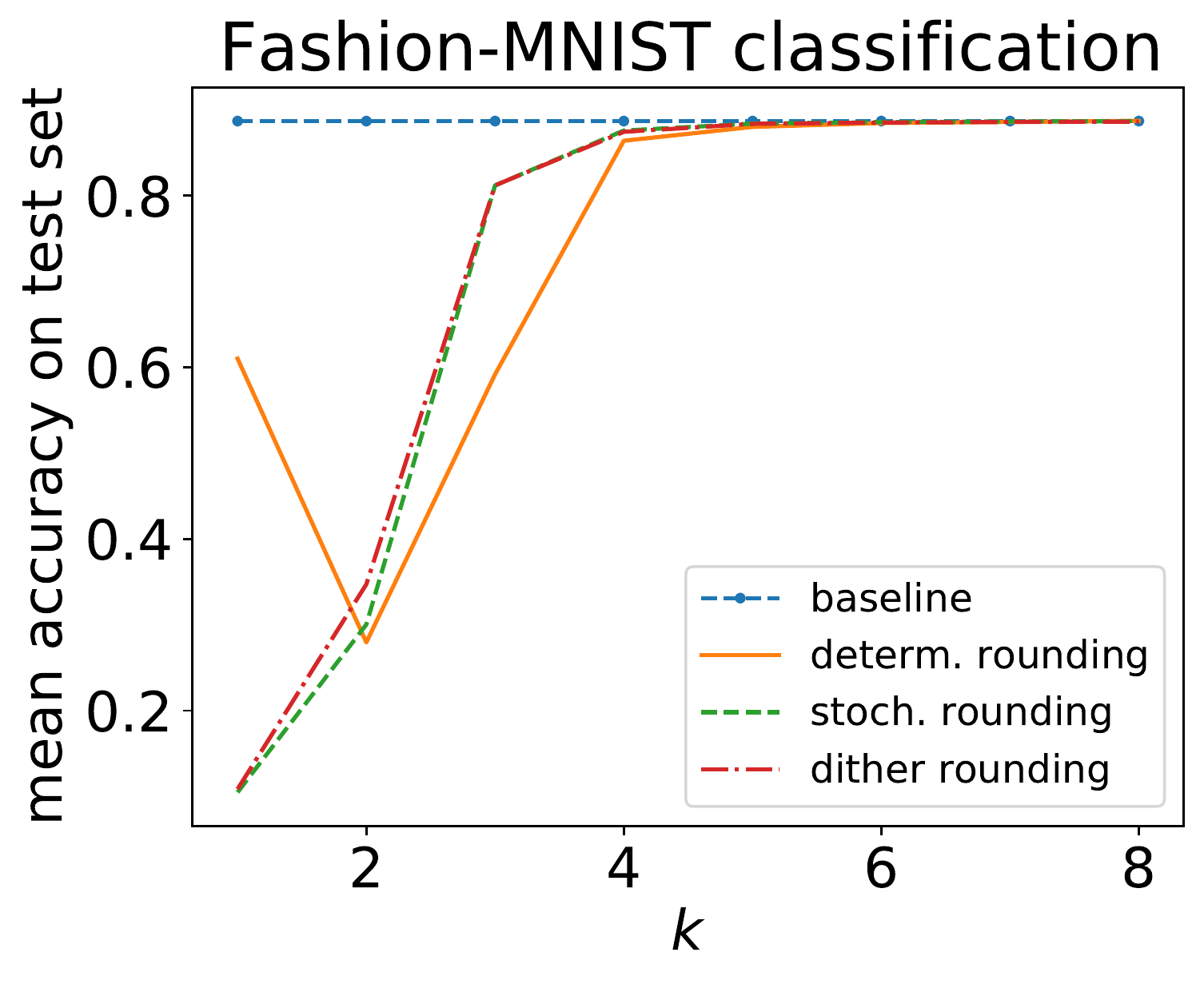}}
\caption{Comparison of deterministic, dither and stochastic rounding for the Fashion MNIST task using a 3-layer MLP. The mean accuracy is shown over $1000$ trials.}
\label{fig:fmnistmlp}
\end{figure}

\begin{figure}[htbp]
\centerline{\includegraphics[width=3in]{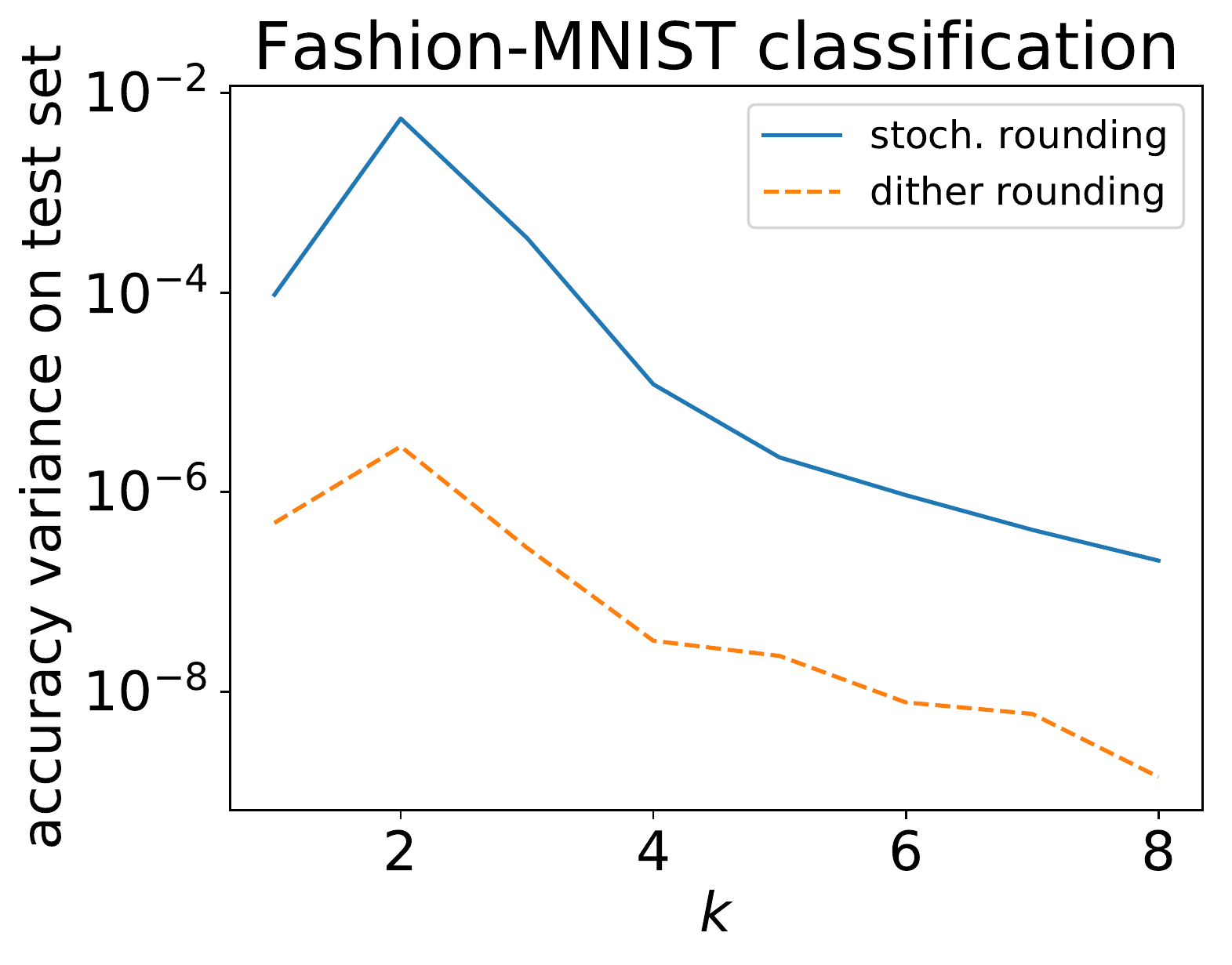}}
\caption{Classification accuracy variance in Fashion MNIST classification using a 3-layer MLP. The sample variance is computed over $1000$ trials.}
\label{fig:fmnistvarmlp}
\end{figure}

We also performed the same experiment for the Fashion MNIST clothing image recognition task \cite{xiao_fashion-mnist:_2017}. Since this is a harder task than MNIST, we use a 3-layer MLP (multi-layer perceptron) network with ReLu activation functions between layers and a softmax output function. Each of the 3 weight matrices, the input data matrix and the intermediate result matrices are rounded separately before applying the matrix multiplication operations. The results are shown in Figs. \ref{fig:fmnistmlp}-\ref{fig:fmnistvarmlp} which illustrate similar trends for this task as well, although the range of $k$ where dither and stochastic rounding are beneficial is much narrower $(3\leq k\leq 4)$.  We plan to provide further results on various variants of dither rounding and explore the various tradeoffs in a future paper.

\section{Conclusions}
We present a hybrid stochastic-deterministic scheme that encompasses the best features of stochastic computing and its deterministic variants by achieving the optimal $\Theta(\frac{1}{N^2})$ asymptotic rate for the EMSE of the deterministic variant while inheriting the zero bias property of stochastic computing schemes with faster convergence to the zero bias. We also show how it can be beneficial in stochastic rounding applications as well, where dither rounding has similar mean performance as stochastic rounding, but with lower variance and is superior to deterministic rounding 
especially in cases where the range of the data is smaller than the full range of the quantizer.

\label{lastpage}


\begin{thebibliography}{10}

\bibitem{gaines:1967}
B.~R. Gaines, ``Stochastic computing,'' in {\em Proc. of the {AFIPS}
  Spring Joint Computer Conference}, pp.~149--156, 1967.

\bibitem{Alaghi2013}
A.~Alaghi and J.~P. Hayes, ``Survey of stochastic computing,'' {\em ACM
  Trans. on Embedded Computing Syst.}, vol.~12, no.~2s, pp.~1--19,
  2013.

\bibitem{Chen2014}
T.-H. Chen and J.~P. Hayes, ``Analyzing and controlling accuracy in stochastic
  circuits,'' in {\em IEEE 32nd International Conference on Computer Design
  (ICCD)}, 2014.

\bibitem{Duarte2015}
R.~P. Duarte, M.~Vestias, and H.~Neto, ``Enhancing stochastic computations via
  process variation,'' in {\em 25th International Conference on Field
  Programmable Logic and Applications (FPL)}, 2015.

\bibitem{Jenson2016}
D.~Jenson and M.~Riedel, ``A deterministic approach to stochastic
  computation,'' in {\em ICCAD}, 2016.

\bibitem{davis:1994}
M.~D. Davis, R.~Sigal, and E.~J. Weyuker., {\em Computability, Complexity, and
  Languages: Fundamentals of Theoretical Computer Science}.
\newblock Academic Press, 1994.

\bibitem{James2013}
G.~James, D.~Witten, T.~Hastie, and R.~Tibshirani, {\em An Introduction to
  Statistical Learning}.
\newblock Springer, 2013.

\bibitem{Hoehfeld1992}
M.~H\"{o}hfeld and S.~E. Fahlman, ``Probabilistic rounding in neural network
  learning with limited precision,'' {\em Neurocomputing}, vol.~4, no.~6,
  pp.~291--299, 1992.

\bibitem{connolly:2020}
M.~P. Connolly, N.~J. Higham, and T.~Mary, ``Stochastic rounding and its
  probabilistic backward error analysis,'' Tech. Rep. MIMS EPrint 2020.12, The
  University of Manchester, 2020.

\bibitem{Gokmen2017}
T.~Gokmen, M.~Onen, and W.~Haensch, ``Training deep convolutional neural
  networks with resistive cross-point devices,'' {\em Frontiers in
  Neuroscience}, vol.~11, 10 2017.

\bibitem{liu:2020}
Y.~Liu, S.~Liu, Y.~Wang, F.~Lombardi, and J.~Han, ``A survey of stochastic
  computing neural networks for machine learning applications,'' {\em IEEE
  Trans. on Neural Networks and Learning Systems}, pp.~1--16, 2020.

\bibitem{Colangelo2018}
P.~Colangelo, N.~Nasiri, E.~Nurvitadhi, A.~K. Mishra, M.~Margala, and
  K.~Nealis, ``Exploration of low numeric precision deep learning inference
  using {I}ntel {FPGA}s,'' in {\em Proc. of the 2018 {ACM/SIGDA}
  International Symposium on Field-Programmable Gate Arrays, {FPGA} 2018,
  Monterey, CA, USA} (J.~H. Anderson and K.~Bazargan,
  eds.), p.~294, 2018.

\bibitem{Choi2019}
J.~Choi, S.~Venkataramani, V.~Srinivasan, K.~Gopalakrishnan, Z.~Wang, and
  P.~Chuang, ``Accurate and efficient 2-bit quantized neural networks,'' in
  {\em Proc. of Machine Learning and Systems 2019, MLSys 2019, Stanford,
  CA, USA} (A.~Talwalkar, V.~Smith, and M.~Zaharia,
  eds.), 2019.

\bibitem{Qin2020}
H.~Qin, R.~Gong, X.~Liu, X.~Bai, J.~Song, and N.~Sebe, ``Binary neural
  networks: {A} survey,'' {\em Pattern Recognition}, vol.~105, p.~107281, 2020.

\bibitem{hopkins:2020}
M.~Hopkins, M.~Mikaitis, D.~R. Lester, and S.~Furber, ``Stochastic rounding and
  reduced-precision fixed-point arithmetic for solving neural ordinary
  differential equations,'' {\em Phisophical Transactions A}, vol.~378,
  p.~20190052, 2020.

\bibitem{Gupta2015a}
S.~Gupta, A.~Agrawal, K.~Gopalakrishnan, and P.~Narayanan, ``Deep learning with
  limited numerical precision,'' in {\em Proc. of the 32nd International
  Conference on Machine Learning}, pp.~1737--1746, Feb. 2015.

\bibitem{strassen:1969}
V.~Strassen, ``Gaussian elimination is not optimal,'' {\em Numer. Math.},
  vol.~13, pp.~354--356, 1969.

\bibitem{Coppersmith1990}
D.~Coppersmith and S.~Winograd, ``Matrix multiplication via arithmetic
  progressions,'' {\em J. Symb. Comput.}, vol.~9, no.~3, pp.~251--280, 1990.

\bibitem{Alman2021}
J.~Alman and V.~V. Williams, ``A refined laser method and faster matrix
  multiplication,'' in {\em Proc. of the 2021 {ACM-SIAM} Symposium on
  Discrete Algorithms, {SODA} 2021}
  (D.~Marx, ed.), pp.~522--539, 2021.

\bibitem{LeCun2010}
Y.~LeCun, C.~Cortes, and C.~Burges, ``{MNIST} handwritten digit database,''
  {\em AT\&T Labs [Online]. Available: http://yann. lecun. com/exdb/mnist},
  vol.~2, 2010.

\bibitem{xiao_fashion-mnist:_2017}
H.~Xiao, K.~Rasul, and R.~Vollgraf, ``Fashion-{MNIST}: a {Novel} {Image}
  {Dataset} for {Benchmarking} {Machine} {Learning} {Algorithms},'' {\em
  arXiv:1708.07747 [cs, stat]}, Aug. 2017.
\newblock arXiv: 1708.07747.

\end{thebibliography}
\end{document}